\documentclass[fleqn,10pt]{wlscirep}
\usepackage[utf8]{inputenc}
\usepackage[T1]{fontenc}

\usepackage{amsmath,amssymb}
\usepackage{graphicx}
\usepackage{caption}
\usepackage{subcaption}
\usepackage{booktabs}
\usepackage{multirow}
\usepackage{url}
\usepackage[dvipsnames]{xcolor}
\usepackage[final]{changes}
\makeatletter
\AddToHook{cmd/added/before}{\def\Changes@AuthorColor{red}}
\AddToHook{cmd/deleted/before}{\def\Changes@AuthorColor{red}}
\AddToHook{cmd/replaced/before}{\def\Changes@AuthorColor{red}}
\makeatother

\title{Realtime forecasting of solar energetic particle event and proton flux using multi-source solar observations and multi-task deep learning}

\author[1]{Yian Yu}
\author[1,*]{Yang Chen}
\author[2]{Lulu Zhao}
\author[3]{Kathryn Whitman}
\author[2]{Ward Manchester}
\author[2]{Tamas Gombosi}

\affil[1]{Department of Statistics, University of Michigan, Ann Arbor, MI 48109, USA}
\affil[2]{Department of Climate and Space Sciences and Engineering, University of Michigan, Ann Arbor, MI 48109, USA}
\affil[3]{NASA Space Radiation Analysis Group, Johnson Space Center, Houston, TX 77058, USA}

\affil[*]{Corresponding author: ychenang@umich.edu}


\begin{abstract}
Solar energetic particle (SEP) events, defined by proton flux exceeding 10 pfu in the $>$ 10 MeV channel, pose major risks to spacecraft operations, astronaut safety, and high-latitude aviation. Due to the complexity and rarity of SEP events, reliable operational SEP forecasting remains an important challenge in space weather. Here we present a novel 24-hour-ahead realtime forecasting framework, \texttt{SEPNET-PRISM}, based on a multi-task learning structure and a thoroughly constructed list of features from multiple sources spanning multiple solar cycles, that jointly predicts SEP event occurrence and future proton and soft X-ray fluxes.  \texttt{SEPNET-PRISM} extends the earlier-introduced \texttt{SEPNET}-based models by integrating a broader range of solar observations, including active-region magnetic parameters from SHARP and SMARP, solar-flare information, coronal mass ejections, soft X-ray flux, and historical $>$ 10 MeV proton flux. As compared with \texttt{SEPNET}, the inclusion of SMARP data expands the temporal coverage of magnetic-field predictors to earlier solar cycles, while flux-based inputs provide additional precursor information.  Evaluation on the CLEAR SEP benchmark dataset shows improved classification performance over the earlier \texttt{SEPNET-O} (operational version of \texttt{SEPNET}) on the newly aligned dataset. The best operational model is obtained when magnetic, radiative, and proton-flux predictors are combined, highlighting the value of expanded historical coverage and complementary precursor information for improving realtime SEP forecasting.
\end{abstract}
\begin{document}

\flushbottom
\maketitle
%
%
\thispagestyle{empty}


\section*{Introduction}
Solar energetic particle (SEP) events are major space weather hazards because enhanced proton fluxes can damage spacecraft systems, increase radiation exposure for astronauts, and disrupt high-latitude aviation operations \cite{Eastwood2017Economic,whitman2024multi}. Operational SEP forecasting remains difficult because SEP occurrence is intermittent and depends on a combination of eruptive activity, magnetic connectivity, and interplanetary transport \cite{REAMES2004,KIM2011,desai2016large,klein2017acceleration}. Reliable forecasting therefore requires both informative observations and models that can extract predictive structure from heterogeneous solar data.

SEP forecasting has traditionally relied on empirical relationships or physics-based simulations. Empirical methods can provide rapid warnings from flares, coronal mass ejections (CMEs), radio bursts, and in situ particle measurements, whereas physics-based approaches aim to model particle acceleration and transport more directly \cite{smart1979pps76,Laurenza2009Atech,Sokolov2004AJ,Luhmann2007295,Hu2017Modeling,Zhao2024SOFIE}. Although both approaches have produced useful forecasting capability, operational performance is often limited by uncertain inputs, incomplete observational coverage, and the trade-off between physical realism and forecast timeliness \cite{erickson1997bruny,Gopalswamy2005JA011158,richardson201425}.

Machine learning has become an increasingly active area in SEP prediction because it can combine diverse solar and heliospheric observations within a unified predictive framework. Recent studies have used flare properties, CME parameters, proton flux, soft X-ray measurements, magnetic-field descriptors, radio observations, and imaging data to support SEP forecasting \cite{Kasapis2022SW,boubrahimi2017prediction,lavasa2021assessing,Dayeh2024SW,ali2025forecasting}. However, recent reviews also highlight substantial variation in data sources, event definitions, forecast windows, and verification metrics, which complicates direct comparison across studies and underscores the need for more harmonized datasets and validation strategies \cite{Whitman20235161}.

In our earlier work, we introduced \texttt{SEPNET}, a multi-task deep learning framework for SEP forecasting that jointly modeled SEP occurrence and related eruptive quantities \cite{yu2026SEPNET}. \added{In \texttt{SEPNET}, multi-task learning incorporates a binary classification head for SEP-event prediction and regression heads for flare and CME properties. The regression outputs are intended to capture broader temporal trajectories and provide auxiliary signals to the classification head. However, these regression tasks are particularly difficult because their target values are highly skewed and strongly correlated with event occurrence.} That study showed that shared representations across related tasks can improve forecasting skill and that magnetic-field measurements provide most important predictive information. However, the earlier framework had two practical limitations. First, magnetic predictors were largely restricted to the SHARP era (year 2010 onward). Second, operationally relevant precursor variables, including continuous GOES soft X-ray (XRSB) flux and historical proton flux, were not fully incorporated into the forecasting pipeline.

Here we address these limitations by utilizing a newly constructed operational SEP forecasting dataset \texttt{SEP-PRISM Data} \cite{Yu2026SEPPRISM} that combines flare information, a reconstructed CME catalog (referred to as CDAWDONKI), GOES XRSB flux, historical $> 10$ MeV proton flux, and a unified magnetic-field product based on SHARP and SMARP observations, hereafter referred to as SMHARP \cite{Bobra2021SMARPSHARP}. By incorporating SMARP data, we extend SHARP-consistent magnetic information into earlier solar cycles. By adding flux-based inputs, we include precursor signals that are directly relevant to operational forecasting. The dataset is designed for 24-hour-ahead prediction of SEP events defined by proton flux exceeding 10 pfu in the $> 10$ MeV channel \cite{whitman2024multi,whitman2026validation}. In total, the constructed operational forecasting dataset contains 14,464 non-overlapping 24-hour samples, of which 650 correspond to positive operational SEP cases in the subsequent 24 hours (approximately 4.5\% positive), highlighting the strong class imbalance arising from the intrinsic rarity of operational SEP events. In contrast to our earlier \texttt{SEPNET} study, which relied on more broadly defined SEP enhancements, the present work defines the positive class directly from operational events exceeding 10 pfu, placing greater emphasis on model performance under realistic class imbalance and operational constraints.

We develop \texttt{SEPNET-PRISM}, an updated operational forecasting model derived from \texttt{SEPNET-O}\cite{yu2026SEPNET} with Proton flux, Radiative, and Integrated Solar Magnetic (PRISM) features. As model inputs, each data source is represented by per-feature summary statistics computed over the preceding non-overlapping 24-hour windows, specifically the minimum, average, and maximum values. This representation compactly captures the recent solar environment and follows our previous study \cite{yu2026SEPNET}, in which summarized window-based predictors proved effective for SEP forecasting. The model then uses sequences of these summarized daily predictor vectors to forecast operational SEP occurrence in the subsequent 24-hour period. The model combines short-term temporal encoding with direct tabular feature modeling and jointly predicts SEP event occurrence, future proton flux, and future XRSB flux. We also evaluate the predictive contribution of different feature groups, including magnetic, flare, CME, X-ray, and proton-flux variables. In summary, with a newly constructed, unified multi-source operational SEP forecasting dataset with extended temporal coverage \cite{Yu2026SEPPRISM}, we develop an enhanced multi-task model for 24-hour-ahead realtime SEP forecasting, quantify feature importance, and assess which combinations of predictors provide the strongest operational forecasting skill.

\section*{Results}
We evaluated \texttt{SEPNET-PRISM} using different predictor groups composed of summarized 24-hour window features derived from various sources of solar activity, including SMHARP magnetic parameters (\texttt{S}), flare-related variables (\texttt{F}), CME-related variables from the aligned CDAWDONKI catalog (\texttt{C}), historical proton flux in the $>10$ MeV channel (\texttt{PF}), and GOES XRSB flux (\texttt{X}), as well as their combinations. Here, \texttt{S\_AR} denotes the subset of SMHARP records matched to flare-associated active regions. We first present the classification and auxiliary performance of \texttt{SEPNET-PRISM} across different feature-group configurations, then compare the updated framework with the earlier \texttt{SEPNET-O} model on the newly aligned dataset, an finally examine feature importance to assess the relative contribution of these candidate predictors. Classification performance is summarized using accuracy (ACC), area under the receiver operating characteristic curve (AUC) \cite{muschelli2020roc}, false positive rate (FPR), F1 score \cite{lipton2014f1}, probability of detection (POD) \cite{wehling2011probability}, false alarm ratio (FAR) \cite{macmillan1985detection}, true skill statistic (TSS) \cite{doswell1990summary}, and Heidke skill score (HSS), following common practice in space weather forecast verification \cite{Leka2019}. Regression performance is summarized using root mean square error (RMSE), mean absolute error (MAE), and coefficient of determination (R$^2$), which is a regression-oriented goodness-of-fit metric.


\subsection*{Realtime SEP classification across feature groups}

\begin{figure}[hpbt]
\centering

\begin{subfigure}{\linewidth}
\centering
\includegraphics[width=\linewidth]{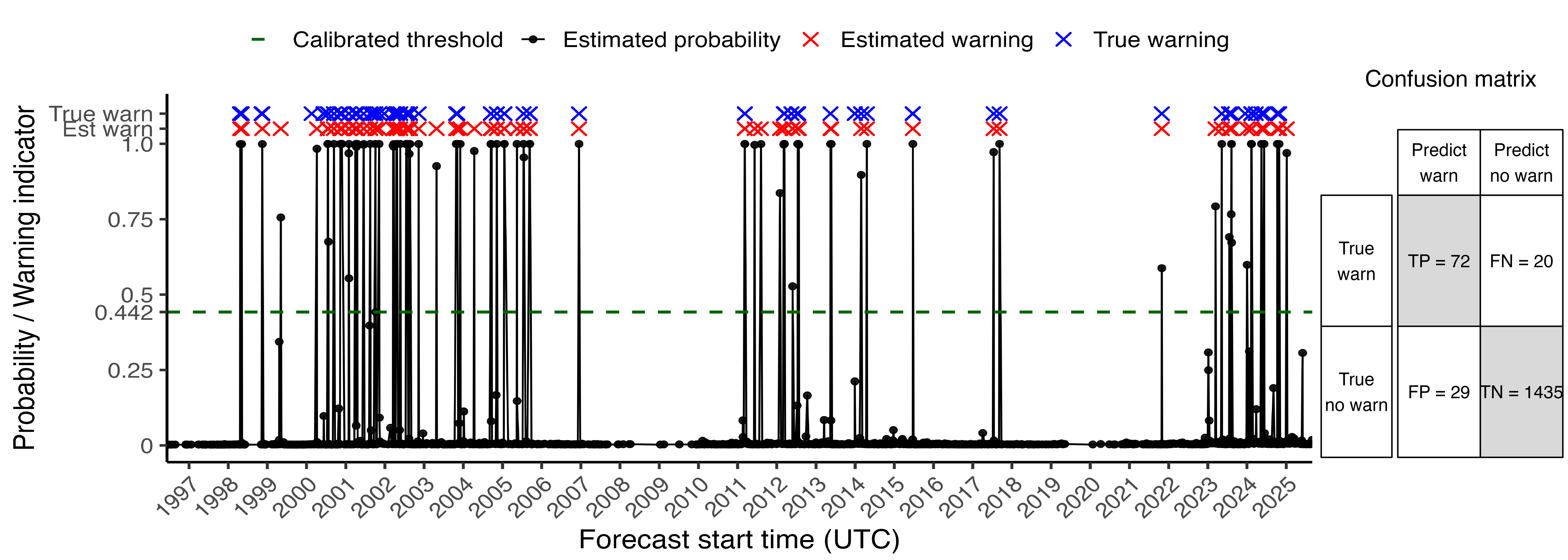}
\caption{Operational SEP classification results for a representative test example.}
\label{fig:case_spf_classi}
\end{subfigure}

\begin{subfigure}{\linewidth}
\centering
\includegraphics[width=\linewidth]{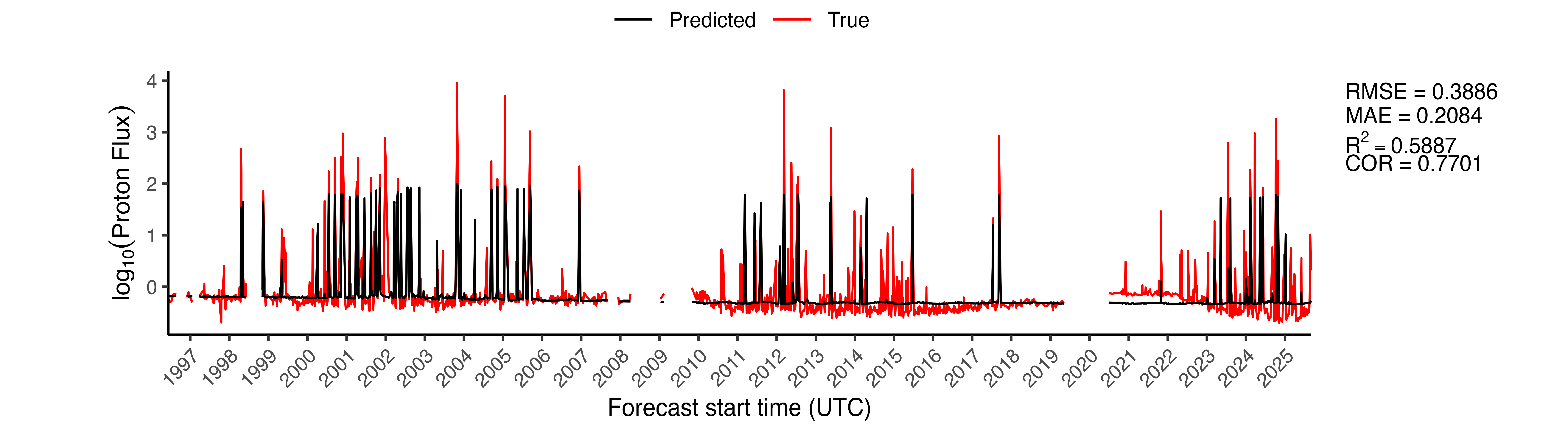}
\caption{Predicted and observed future maximum logarithmic proton flux in the $> 10$ MeV channel.}
\label{fig:case_spf_reg_pf}
\end{subfigure}

\begin{subfigure}{\linewidth}
\centering
\includegraphics[width=\linewidth]{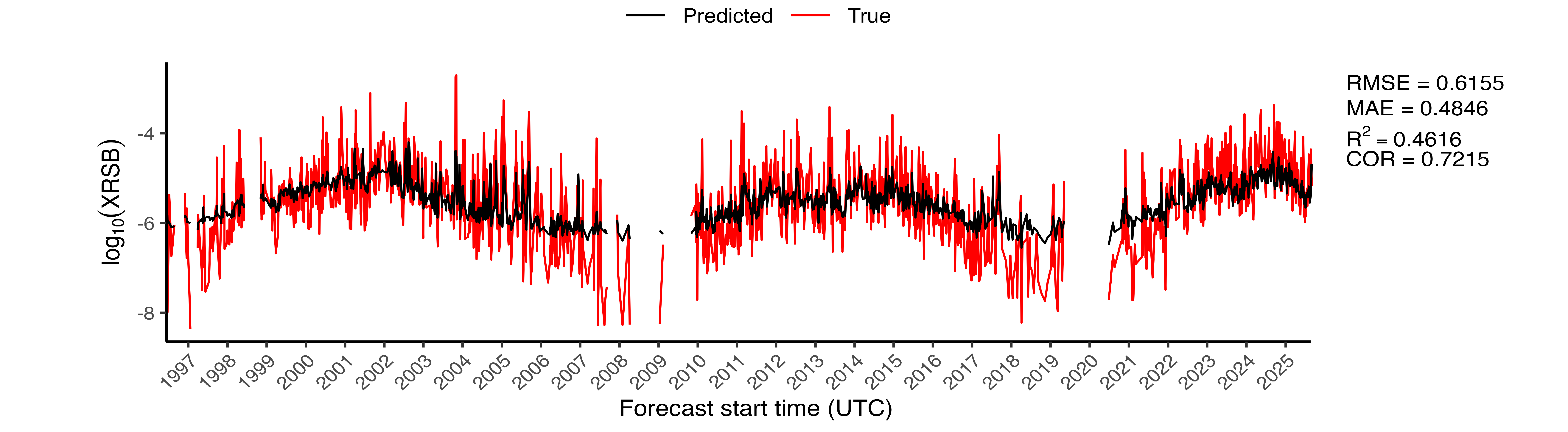}
\caption{Predicted and observed future maximum logarithmic XRSB flux.}
\label{fig:case_spf_reg_XRSB}
\end{subfigure}
\caption{Representative test-set predictions of \texttt{SEPNET-PRISM} for the \texttt{S+PF+F} feature configuration under one random seed.}
\label{fig:case_spf}
\end{figure}

We first evaluated \texttt{SEPNET-PRISM} across different feature configurations. Representative predictions for the \texttt{S+PF+F} configuration are shown in Figure \ref{fig:case_spf_classi}, and the full set of classification metrics is summarized in Table \ref{tab:classifi_results}. Among the single-group models, the SMHARP magnetic feature set provided the strongest overall classification performance. This result is consistent with the earlier \texttt{SEPNET} study, in which SHARP parameters showed strong SEP forecasting skill. By contrast, the restricted \texttt{S\_AR} feature set performed less well than the full SMHARP set, suggesting that broader magnetic-field context retains useful predictive information beyond flare-associated active regions alone. Among the non-magnetic single-group models, historical proton flux performed best. Flare-only predictors provided moderate skill, whereas CME-only and XRSB-only models were weaker, especially in terms of F1 score and skill scores. 

Combining complementary feature groups substantially improved performance. Relative to the proton-flux-only model, adding flare information or XRSB improved classification skill, with \texttt{PF+F} outperforming \texttt{PF+X} and suggesting that flare-related predictors provide additional information that is not fully captured by XRSB flux alone. As in the original \texttt{SEPNET} framework, flare-related predictors were constructed from flare duration, rise time, and the logarithm of flare peak flux, whereas GOES flare classes are defined from soft X-ray emission, the flare parameters offer complementary temporal context beyond that contained in XRSB. 

The strongest classification performance overall was achieved by models that combined SMHARP magnetic parameters with proton-flux and radiative proxies. In particular, \texttt{S+PF+X} yielded the highest AUC, \texttt{S+PF+F} achieved the highest F1 score and HSS, and \texttt{S+PF+X+F} produced the highest TSS, all with consistently strong performance across the remaining metrics. By contrast, augmenting these best-performing feature combinations with CME-related predictors did not systematically improve performance and in some cases slightly degraded AUC or increased FAR. 

\begin{table}[hpbt!]
\centering
\caption{Classification performance of \texttt{SEPNET-PRISM} for operational SEP forecasting under different input-feature configurations. Feature abbreviations are as follows: \texttt{S}, SMHARP magnetic parameters; \texttt{S\_AR}, SMHARP parameters restricted to flare-associated active regions; \texttt{PF}, historical proton flux in the $>10$ MeV channel; \texttt{X}, GOES soft X-ray (XRSB) flux; \texttt{F}, flare-related variables; and \texttt{C}, CME-related variables from the aligned CDAWDONKI catalog. Metric abbreviations are as follows: ACC, accuracy; AUC, area under the receiver operating characteristic curve; FPR, false positive rate; POD, probability of detection; FAR, false alarm ratio; TSS, true skill statistic; and HSS, Heidke skill score. Results are the median test-set metrics over five independent runs.}
\label{tab:classifi_results}
\begin{tabular}{lcccccccc}
\toprule
Features & ACC & AUC & FPR & F1 & POD & FAR & TSS & HSS \\
\midrule
\texttt{PF} & 0.9297 & 0.8454 & 0.0522 & 0.4129 & 0.5461 & 0.6607 & 0.4939 & 0.3785 \\
\texttt{X} & 0.8991 & 0.6754 & 0.0735 & 0.2215 & 0.3154 & 0.8186 & 0.2419 & 0.1745 \\
\texttt{F} & 0.9062 & 0.8305 & 0.0686 & 0.3773 & 0.4946 & 0.6931 & 0.4260 & 0.3297 \\
\texttt{C} & 0.9165 & 0.6632 & 0.0392 & 0.2832 & 0.2602 & 0.6878 & 0.2211 & 0.2393 \\
\texttt{PF+X} & 0.9517 & 0.8925 & 0.0334 & 0.5427 & 0.6369 & 0.5269 & 0.6035 & 0.5179 \\
\texttt{PF+F} & 0.9577 & 0.9154 & 0.0265 & 0.6573 & 0.7008 & 0.3801 & 0.6743 & 0.6349 \\
\texttt{PF+X+F} & 0.9576 & 0.9179 & 0.0260 & 0.6548 & 0.6915 & 0.3757 & 0.6655 & 0.6323 \\
\texttt{S} & 0.9523 & 0.9040 & 0.0285 & 0.5410 & 0.5761 & 0.4895 & 0.5476 & 0.5160 \\
\texttt{S\_AR} & 0.9410 & 0.8694 & 0.0327 & 0.5235 & 0.5337 & 0.4862 & 0.5010 & 0.4920 \\
\texttt{S+F} & 0.9480 & 0.9072 & 0.0348 & 0.5316 & 0.6109 & 0.5278 & 0.5761 & 0.5046 \\
\texttt{S+C} & 0.9397 & 0.8743 & 0.0370 & 0.5715 & 0.6096 & 0.4614 & 0.5726 & 0.5392 \\
\texttt{S+PF+X} & 0.9688 & 0.9253 & 0.0166 & 0.6806 & 0.6826 & 0.3206 & 0.6660 & 0.6642 \\
\texttt{S+PF+F} & 0.9658 & 0.9080 & 0.0167 & 0.7035 & 0.6869 & 0.2743 & 0.6703 & 0.6854 \\
\texttt{S+PF+X+F} & 0.9639 & 0.9161 & 0.0194 & 0.6956 & 0.6978 & 0.3027 & 0.6784 & 0.6764 \\
\texttt{S+PF+X+F+C} & 0.9573 & 0.9086 & 0.0220 & 0.7014 & 0.6916 & 0.2837 & 0.6696 & 0.6785 \\
\texttt{S+PF+F+C} & 0.9573 & 0.9065 & 0.0222 & 0.7018 & 0.6940 & 0.2889 & 0.6718 & 0.6789 \\
\bottomrule
\end{tabular}
\end{table}

\subsection*{Regression performance for auxiliary targets}

Because \texttt{SEPNET-PRISM} was trained in a multi-task setting, we also evaluated regression performance for future maximum logarithmic proton flux in the $>$ 10 MeV channel and future maximum logarithmic XRSB flux over the subsequent 24-hour window. Representative regression predictions for the \texttt{S+PF+F} configuration are shown in Figure \ref{fig:case_spf}\subref{fig:case_spf_reg_pf},\subref{fig:case_spf_reg_XRSB}, and summary metrics are reported in Table \ref{tab:regress_results}. Overall, the strongest regression performance was again obtained when magnetic, proton-flux, and radiative predictors were combined. These results indicate that the auxiliary tasks are not merely secondary outputs, but help the shared representation encode aspects of eruptive intensity and precursor behavior that are also relevant to SEP event classification. 

For future proton flux, \texttt{S+PF+X} yielded the lowest RMSE and MAE, whereas \texttt{S+PF+X+F} and \texttt{S+PF+F} achieved the strongest correlations. For future XRSB flux, the best results were obtained from feature sets centered on flare, X-ray, and magnetic information, which is physically plausible because future soft X-ray behavior is expected to be more closely related to flare- and radiative-activity indicators.

\begin{table}[hpbt!]
\centering
\caption{Regression performance of \texttt{SEPNET-PRISM} for future maximum logarithmic proton flux in the $>10$ MeV channel and future maximum logarithmic XRSB flux over the subsequent 24-hour window under different input-feature configurations. Feature abbreviations are the same as in Table \ref{tab:classifi_results}. RMSE denotes root mean square error, MAE denotes mean absolute error, and R$^2$ denotes the coefficient of determination. Results are the median test-set metrics over five independent runs.}
\label{tab:regress_results}
\begin{tabular}{lcccccc}
\toprule
\multirow{2}{*}{Features} & \multicolumn{3}{c}{$\log_{10}$(Proton Flux)} & \multicolumn{3}{c}{$\log_{10}$(XRSB)} \\
\cmidrule(lr){2-4}\cmidrule(lr){5-7}
& RMSE & MAE & R$^2$ & RMSE & MAE & R$^2$ \\
\midrule
\texttt{PF} & 0.7011 & 0.3684 & -0.5048 & 1.0547 & 0.8409 & 0.2712 \\
\texttt{X} & 0.7377 & 0.3801 & -0.6475 & 0.7582 & 0.5742 & 0.6235 \\
\texttt{F} & 0.6853 & 0.3567 & -0.2052 & 0.8364 & 0.6563 & 0.1240 \\
\texttt{C} & 0.5897 & 0.2967 & 0.0661 & 0.9669 & 0.7381 & 0.0928 \\
\texttt{PF+X} & 0.5278 & 0.2579 & 0.1570 & 0.7402 & 0.5524 & 0.6412 \\
\texttt{PF+F} & 0.4771 & 0.2529 & 0.4156 & 0.8534 & 0.6734 & 0.0882 \\
\texttt{PF+X+F} & 0.4782 & 0.2512 & 0.4127 & 0.7127 & 0.5441 & 0.3636 \\
\texttt{S} & 0.4555 & 0.2202 & 0.3351 & 0.7888 & 0.6219 & 0.4451 \\
\texttt{S\_AR} & 0.5210 & 0.2620 & 0.2668 & 0.7654 & 0.5936 & 0.1208 \\
\texttt{S+F} & 0.4652 & 0.2249 & 0.3064 & 0.6649 & 0.5267 & 0.6078 \\
\texttt{S+C} & 0.5098 & 0.2630 & 0.3399 & 0.8265 & 0.6344 & 0.1848 \\
\texttt{S+PF+X} & 0.3884 & 0.1967 & 0.5162 & 0.7587 & 0.5942 & 0.4898 \\
\texttt{S+PF+F} & 0.4108 & 0.2189 & 0.5370 & 0.6209 & 0.4806 & 0.4394 \\
\texttt{S+PF+X+F} & 0.4081 & 0.2168 & 0.5430 & 0.6249 & 0.4824 & 0.4305 \\
\texttt{S+PF+F+C} & 0.4490 & 0.2398 & 0.5203 & 0.6214 & 0.4819 & 0.3620 \\
\texttt{S+PF+X+F+C} & 0.4549 & 0.2431 & 0.5075 & 0.7100 & 0.5515 & 0.1696 \\
\bottomrule
\end{tabular}
\end{table}

\subsection*{Comparison with the earlier \texttt{SEPNET-O} framework}

\begin{figure}[hpbt]
\centering
\includegraphics[width=\linewidth]{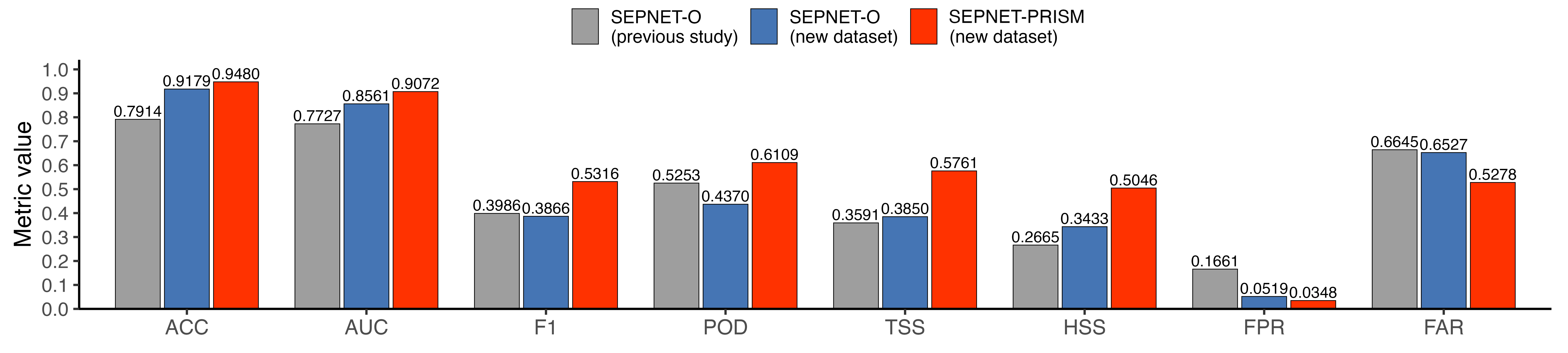}
\caption{Comparison of classification performance among the earlier \texttt{SEPNET-O} results reported previously, \texttt{SEPNET-O} evaluated on the newly aligned dataset, and \texttt{SEPNET-PRISM} under the \texttt{S+F} feature configuration.}
\label{fig:sepnets_comparison_bar}
\end{figure}

To separate the effect of dataset preparation from that of model development, we also evaluated the earlier \texttt{SEPNET-O} model on the newly aligned dataset using the magnetic and flare feature set adopted in the previous study \cite{yu2026SEPNET}. The comparison is summarized in Figure \ref{fig:sepnets_comparison_bar}. On the newly prepared dataset, \texttt{SEPNET-O} showed clear improvement relative to its earlier reported performance, indicating that the revised data construction and expanded historical coverage provide a stronger basis for operational SEP forecasting. \texttt{SEPNET-PRISM} then provided further gains over \texttt{SEPNET-O}, showing that both the improved data construction and the updated model architecture contributed to the final forecasting performance. \added{Besides this, we also compared the test results on the SEPVAL dataset \cite{whitmanSEPVAL}; the detailed results are provided in the Supplementary Information and further demonstrate the improved performance of \texttt{SEPNET-PRISM}.}

This comparison indicates that progress in operational SEP forecasting in the present framework arose from complementary factors: a richer and better aligned multi-source dataset, and a model design that more effectively exploits temporal context, nonlinear feature interactions, and physically meaningful auxiliary tasks. 

\subsection*{Feature importance of candidate predictors}

\begin{figure}[pbht!]
\centering
\includegraphics[width=\linewidth]{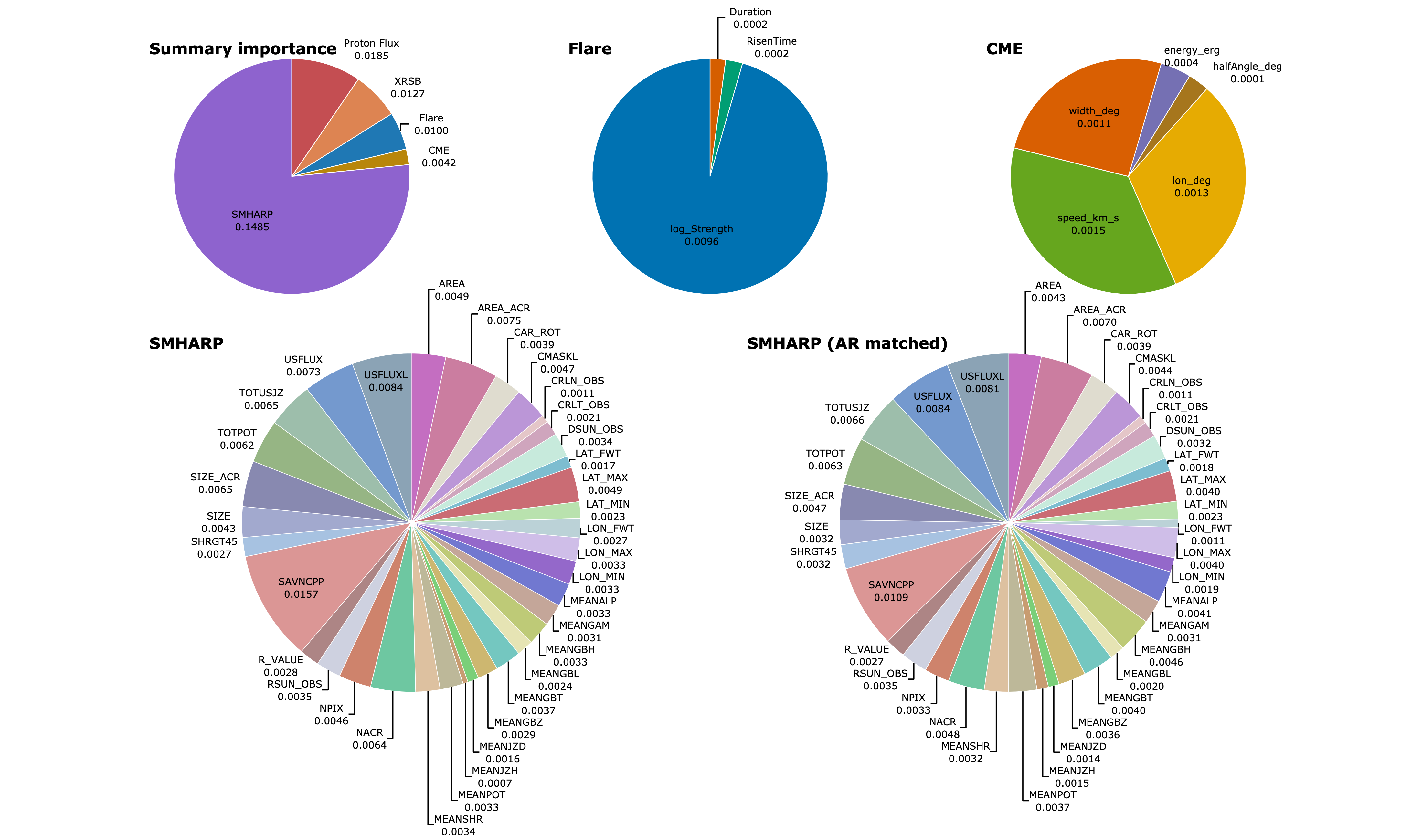}
\caption{Relative feature importance of predictor groups for operational SEP forecasting derived from permutation importance in repeated, class-balanced random forest models. The pie charts provide a compact summary of the aggregate importance of predictors by data source and variable family. For each predictor, importance was aggregated across the minimum, average, and maximum values computed over the preceding non-overlapping 24-hour window. The corresponding predictor-level bar plots are provided in the Supplementary Information.}
\label{fig:feature_importance}
\end{figure}

To help interpret the forecasting results, we examined the predictive contribution of candidate variables using repeated class-balanced random forest models with permutation importance. The aggregated results are shown in Figure \ref{fig:feature_importance}. Magnetic parameters from the SMHARP data showed the strongest and most consistent importance across aggregation statistics. Proton flux in the $> 10$ MeV channel and GOES XRSB flux also showed appreciable importance, suggesting that direct precursor flux information are highly informative for 24-hour-ahead operational SEP forecasting. Flare-related variables made more moderate contributions, with logarithmic intensity measurements appeared to be informative, whereas CME-related variables from the CDAWDONKI catalog had weaker impact. These importance patterns aligned with the performance experiments. The strongest classification and regression results were obtained with feature sets centered on magnetic, proton-flux, and radiative predictors, whereas CME variables contributed less consistently to the strongest-performing configurations.

\section*{Discussion}

This study shows that operational SEP forecasting benefits from combining complementary information sources rather than relying on any single class of predictors. Across the tested configurations, the strongest performance was obtained when magnetic-field information was combined with historical proton flux and radiative indicators of eruptive activity. These results suggest that SEP forecasting skill depends not only on model architecture, but also on the breadth, temporal coverage, and physical relevance of the underlying dataset. 

A central result is that historical proton flux, flare and X-ray activity, and magnetic-field information provide the most informative predictors for 24-hour-ahead SEP forecasting. Feature importance analysis identified proton flux and XRSB variables as dominant contributors, with magnetic parameters adding valuable large-scale active-region context. By contrast, CME-related variables showed weaker and less consistent contributions in the present framework. \replaced{This behavior may result from several factors, including the limitation of a fixed non-overlapping 24-hour windowing strategy in training data construction, the short time gap between CME occurrence and SEP onset, and possible redundancies among CME-related, magnetic, and other eruptive predictors \cite{Li_2026,Kazachenko_2017}. In other words, while CME properties are physically important for SEP production, their incremental value in this particular predictive setting may be reduced by the way the data are summarized and aligned for forecasting. Hourly near-real-time forecasting remains an active direction in our ongoing and future work.}{This pattern was supported by the model-comparison experiments, in which the best classification and regression performance was achieved by combinations centered on magnetic, proton-flux, and radiative inputs rather than by models including CME features}. 

The results also clarify the role of multi-task learning in this setting. Joint predictions of future proton flux and future XRSB flux do not simply add auxiliary outputs; it encourages the shared representation to capture information related to eruptive intensity and precursor structure, which is also relevant to SEP classification. This is consistent with the improved regression performance of combined feature sets and with the stronger classification skill obtained by the best multi-source models. \added{At the same time, although the auxiliary regression tasks improve the model’s ability to reproduce the overall proton flux trajectory, the exact peak flux value remains more difficult to estimate for strong events with highly-skewed flux distributions.}

From an operational \added{or near-real-time forecasting} perspective, the present framework targets a practically relevant 24-hour warning horizon and achieves strong skill while maintaining low false-positive rates in the best-performing configurations. At the same time, false alarms remain an important limitation for downstream use, particularly in real-time warning services where excessive alerts can erode user trust and reduce actionability. \added{In addition, the availability of near-real-time input features and whether they provide sufficient lead time for SEP forecasting remain open questions.} The current results should therefore be viewed as an important step toward operational deployment rather than a final real-time forecasting solution. 

Several limitations also define the next stages of model development. First, the present experiments use simple random splits, whereas future operational evaluation should emphasize temporally ordered testing to better mimic hourly real-time forecasting conditions. Second, permutation importance provides a useful global ranking of predictors but does not yet explain why a warning is issued at a specific forecast time. Extending this framework to hourly real-time forecasting, reducing false warnings, and incorporating time-resolved explanation methods such as \replaced{SHapley Additive exPlanations (SHAP) \cite{lundberg2017shap,Daniel2022GBM}}{SHAP} are promising next steps. 

\section*{Methods}

\subsection*{Dataset construction and preprocessing}

The primary prediction target in this study was the occurrence of an operational SEP event within the subsequent 24-hour prediction window, where an event was defined by proton flux exceeding 10 pfu in the $>10$ MeV channel. These events were obtained from the CLEAR SEP benchmark dataset, provided by the \texttt{FetchSEP} Python module \cite{fetchsep}, spanning 3 February 1986 to 10 September 2025 and containing 267 operational SEP events. To support feature importance analysis and predictive modeling, \texttt{SEP-PRISM Data}\cite{Yu2026SEPPRISM} integrates observations from multiple solar missions and event catalogs, including flare properties, active-region magnetic parameters, CME records, GOES XRSB flux, and historical proton flux. The temporal coverage of all data sources is summarized therein. \added{We provide a brief description of the \texttt{SEP-PRISM Data} below.}

Flare events were retrieved from the Heliophysics Event Knowledgebase through SunPy's Fido interface \cite{Barnes2020SunPy,hek_dataset}, using GOES as the observing source. The resulting flare catalog contains 78,724 records over 3 February 1986 to 10 September 2025. For each flare, we extracted the start, peak, and end times, GOES class, NOAA active-region number, and heliographic coordinates, and derived flare duration, rise time, and logarithmic peak strength. 

Magnetic-field predictors were obtained from the SDO/HMI SHARP and SOHO/MDI SMARP data products through the Joint Science Operations Center using the \texttt{drms} client \cite{Glogowski2019}. SHARP data are available at 12-min cadence, whereas SMARP data are available at 96-min cadence. Because the SHARP archive begins in May 2010, we incorporated SMARP data, which cover 23 April 1996 to 27 October 2010, to extend SHARP-like magnetic information back to 1996 through a feature-wise alignment procedure based on overlapping active-region observations\cite{Yu2026SEPPRISM}. The resulting aligned dataset, referred to as SMHARP, provides a unified magnetic predictor archive with extended temporal coverage. 

Although SHARP and SMARP differ in instrumental origin and available variables, they share a subset of common or physically related parameters. Previous studies have attempted to combine SHARP and SMARP using global normalization or regression-based rescaling, focusing primarily on parameters that are shared between the two datasets \cite{Sun2022flare,Kosovich2024SMHARP}. In this work, we adopt a feature-wise alignment framework that leverages the full set of matched active regions during the overlapping period from May to October 2010. The alignment is performed at the active-region level, using identifiers when available and spatial proximity otherwise. Each SHARP parameter is reconstructed using multivariate regression models based on SMARP features, allowing parameter-specific correction beyond simple linear rescaling. In addition, a two-stage modeling strategy is employed, in which shared parameters are first aligned and subsequently used to inform the reconstruction of SHARP-specific parameters. Details of the alignment procedure and validation can be found in the \texttt{SEP-PRISM Data} descriptor\cite{Yu2026SEPPRISM}.

This procedure improved agreement between SMARP-derived estimates and SHARP observations, with higher correlations and lower errors across most parameters. For example, Pearson correlations for \texttt{USFLUXL}, \texttt{MEANGBL}, and \texttt{R\_VALUE} increased to 0.9849, 0.8433, and 0.9007, while geometric parameters maintained near-perfectly consistent. Compared with existing rescaling-based methods, the proposed framework provides a more flexible and physically consistent alignment, enabling reliable reconstruction of SHARP-like parameters prior to 2010. The resulting SMHARP dataset extends SHARP-consistent time series back to 1996 and provides a unified predictor set for long-term space weather modeling.

In addition to the full SMHARP archive, we construct an AR-matched subset to isolate magnetic measurements associated with flare-related active regions. For flare records with identified NOAA active-region numbers, we match the corresponding SMHARP records using the active-region identifier when available and spatial proximity otherwise. This AR-matched SMHARP subset provides localized magnetic information tied to flare-producing regions, while the full SMHARP archive retains broader contextual information on active-region magnetic conditions over time.

The CME catalogs retrieved from NASA’s DONKI database \cite{ccmc_donki} are temporally limited to events that occurred from 2010 onward. To obtain historical information on CME events prior to this period, we incorporated the CDAW CME catalog derived from SOHO/LASCO observations \cite{gopalswamy2025CDAW}, which extends the temporal coverage back to 1996. For these two CME catalogs, we encountered the same issue of feature misalignment as in the SHARP and SMARP datasets. After removing all low–quality event entries from the CDAW catalog, we constructed a linear mapping between the two datasets, following a similar procedure used for the SMARP and SHARP data products. Unlike the continuously sampled SHARP parameters, the CME catalogs consist of discrete events. To align the CME catalogs, we adopted the CDAW list as the reference catalog. For each event in CDAW, we searched for the closest corresponding event in DONKI; if the temporal difference between the two events exceeded 24 hours, the event from CDAW list was considered unmatched and excluded from the mapping procedure. \added{This 24-hour tolerance was used only as a practical matching criterion to account for catalog timing differences.} After performing this matching step and discarding all unmatched CDAW events, we constructed the linear mapping between the CDAW and DONKI CME parameters analogously to the method applied to SHARP and SMARP. The resulting catelog, denoted CDAWDONKI, contains 17,463 records and extends DONKI-like CME information back to 1996. The detailed procedure is described in the \texttt{SEP-PRISM Data} descriptor\cite{Yu2026SEPPRISM}.

GOES X-ray Sensor data in the 1--8 $\mathring{A}$ channel (XRSB) were retrieved across multiple GOES satellites, with recent data gaps supplemented by the NOAA Space Weather Prediction Center feed. Proton fluxes in the $> 10$ MeV channel were obtained from NASA's ISWA HAPI service using primary and secondary GOES proton products, both sampled at a 5-minute cadence.

\begin{figure}[hpbt]
\centering
\includegraphics[width=\linewidth]{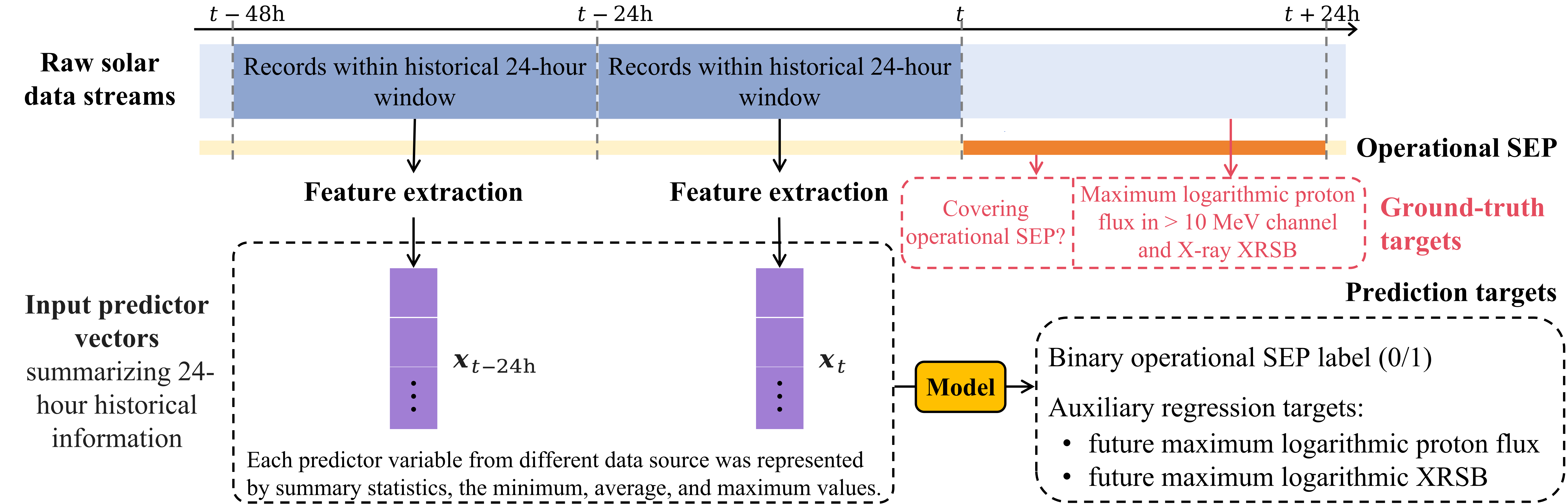}
\caption{Schematic of the preprocessing workflow used to convert heterogeneous solar observations into supervised-learning samples. Data from each source were aggregated into fixed, non-overlapping 24-hour historical windows, and each predictor variable was represented by its minimum, mean, and maximum values within the window. Prediction targets were defined over the subsequent 24-hour window and included a binary operational SEP label together with two auxiliary regression targets.}
\label{fig:Data_Preprocessing}
\end{figure}

For supervised learning, all heterogeneous observations were aggregated into fixed, non-overlapping 24-hour windows. This preprocessing and prediction-target construction is illustrated schematically in Figure \ref{fig:Data_Preprocessing}. Within each window, each predictor variable was represented by three summary statistics, the minimum, average, and maximum values. Thus, the predictor groups analyzed in this study (\texttt{S}, \texttt{S\_AR}, \texttt{F}, \texttt{C}, \texttt{PF}, and \texttt{X}) refer to collections of summarized window-level features rather than raw time series or event lists. This representation provides a common input format across heterogeneous data sources while retaining information about the recent solar environment over the preceding 24 hours, and follows the summarized-window strategy used in our previous study. The binary prediction target was defined over the subsequent 24-hour interval as whether an SEP event occurred. In addition, two auxiliary regression targets were defined as the future maximum logarithmic proton flux in the $> 10$ MeV channel and the future maximum logarithmic XRSB flux. The resulting dataset comprised 14,464 samples, of which 650 were labeled as positive operational SEP cases in the subsequent 24-hour window. A sample was labeled positive if the future 24-hour window overlapped in time with any operational SEP event list in CLEAR SEP benchmark dataset, defined by proton flux exceeding 10 pfu in the $>$ 10 MeV channel from event start to end. In contrast to the earlier \texttt{SEPNET} study \cite{yu2026SEPNET}, where future operational SEP events were inferred from more inclusive general SEP events \replaced{defined based on enhancements above the GOES background level (indicated by an arbitrarily low threshold of 10$^{-6}$ pfu) in the $>$ 10 MeV channel}{exceeding 10$^{-6}$ pfu in the $>$ 10 MeV channel}, the present work labels positives directly from operational events, leading to a smaller positive fraction that more accurately reflects operational forecasting conditions. 

\subsection*{Feature importance analysis}

As a preliminary exploratory step, we assessed feature importance using repeated, class-balanced random forest models. All candidate predictors were first scaled to the range $[0,1]$. Missing values were handled through a two-stage procedure consisting of $k$-nearest-neighbour imputation with $k=5$, followed by median imputation for any remaining missing entries. To account for class imbalance and sampling variability, we used repeated stratified resampling equivalent to five-fold cross-validation repeated five times. Within each split, a random forest with 1000 trees was trained using class-balanced bootstrap samples. 

Feature importance was quantified using permutation importance, defined as the reduction in predictive performance when a feature is randomly permuted. This analysis was used to guide the subsequent model experiments by identifying candidate variables and feature groups that were likely to be informative for operational SEP forecasting. 

\subsection*{SEPNET-PRISM architecture}

\texttt{SEPNET-PRISM} extends the earlier \texttt{SEPNET-O} framework by incorporating short-term temporal evolution through sequences of summarized predictor windows. For each reference time $t$, we constructed a sequence of length $L=2$, consisting of the summarized predictor vectors from the current and previous 24-hour windows. This yields 48 hours of recent history while preserving a compact tabular representation of heterogeneous inputs. Missing historical steps were padded with zeros. 

\begin{figure}[hpbt]
\centering
\includegraphics[width=\linewidth]{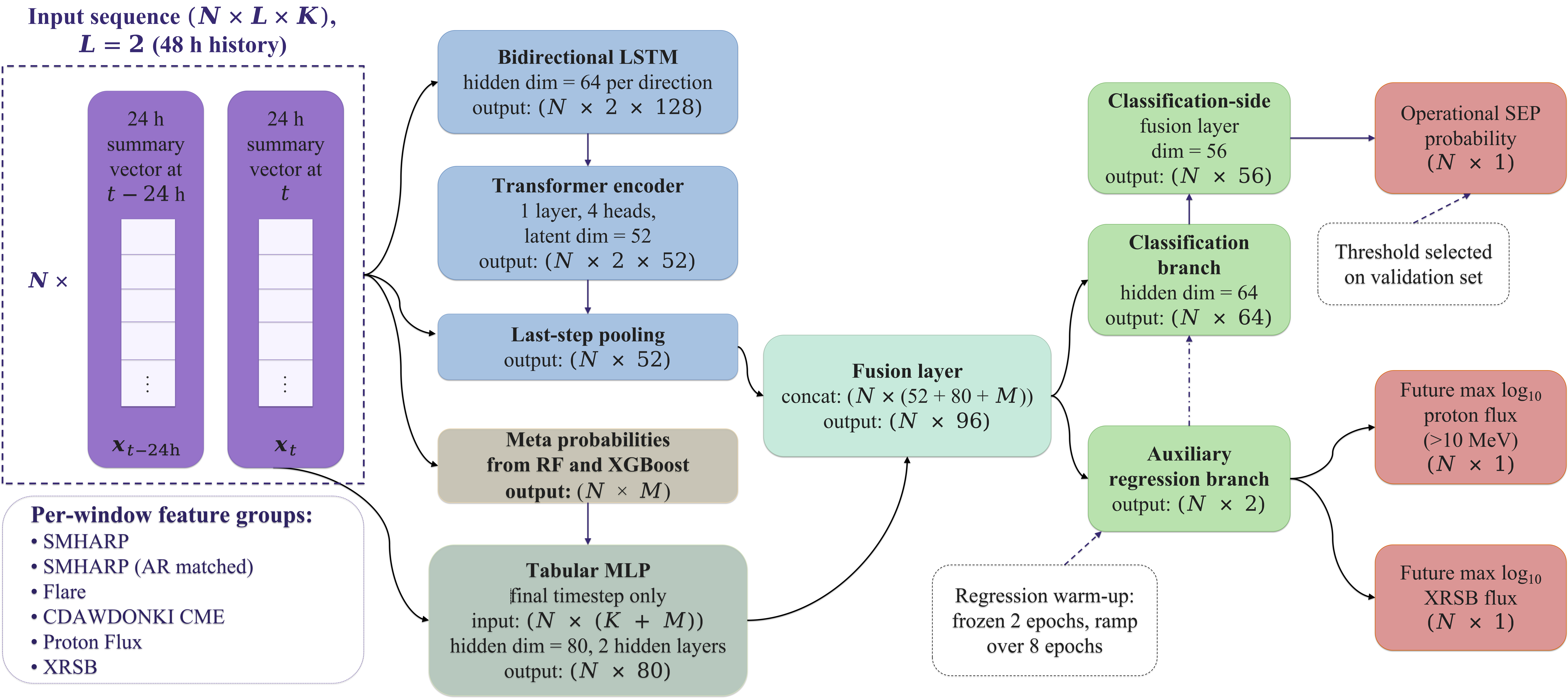}
\caption{Architecture of \texttt{SEPNET-PRISM}. Each sample is represented as a sequence of two summarized 24-hour predictor vectors. A temporal branch processes the sequence using a bidirectional LSTM followed by a Transformer encoder, while a parallel tabular branch encodes the most recent predictor vector. The resulting representations are fused and passed to task-specific heads for operational SEP classification and auxiliary regression of future proton and XRSB fluxes.  The framework also incorporates validation-based threshold calibration and regression warm-up during training.}
\label{fig:sepnets_ov2_architecture}
\end{figure}

The model uses a hybrid architecture that combines a temporal branch and a tabular branch. The temporal branch processes the two-step sequence with a bidirectional LSTM encoder followed by a Transformer encoder and last-step pooling. In parallel, the most recent summarized predictor vector is processed by a multilayer perceptron. The resulting latent representations are concatenated and mapped into a shared representation, which is then passed to task-specific heads for binary SEP classification and auxiliary regression of future proton and XRSB fluxes. Meta-probabilities from external models, such as random forest and XGBoost, are also be incorporated. A schematic overview of the proposed \texttt{SEPNET-PRISM} architecture is shown in Figure \ref{fig:sepnets_ov2_architecture}.

The overall loss function combines regression and classification objectives, including weighted binary cross-entropy and focal loss for the classification branch and a regression loss for the continuous auxiliary targets. To improve training stability, the regression branch is gradually introduced during an initial warm-up stage. Balanced mini-batch sampling, positive-class weighting, and focal loss are used to mitigate class imbalance and improve sensitivity to rare SEP events. 

\subsection*{Training, validation, and evaluation}

The dataset was divided into training, validation, and testing sets in a supervised-learning framework. In the present random i.i.d. implementation, 20\% of the available samples were reserved for testing, and 25\% of the remaining training pool was used for validation. The validation set was used for model selection and threshold calibration. Rather than using the default threshold of 0.5, the final warning threshold was selected by scanning candidate probability cutoffs on the validation set and choosing the value that optimized a skill-based criterion involving HSS and TSS, i.e., $\min(\mathrm{TSS}, \mathrm{HSS})$. 

Hyperparameters were tuned using Optuna over a search space covering optimization parameters, regularization, class weighting, focal-loss coefficients, and relative loss weights between the regression and classification branches. For each feature configuration, experiments were repeated over five random seeds, and reported classification and regression metrics summarize median test-set performance across these runs. 

\bibliography{reference}

\section*{Acknowledgments}
The authors thank Ian Richardson (University of Maryland, NASA GSFC) and A. Steve Johnson (Leidos, NASA JSC SRAG) for their efforts in acquiring and preparing the dataset used in this study. The authors thank Spiridon Kasapis for discussions on the SMARP data.

\section*{Funding}

This work is supported by the NASA Space Weather Center of Excellence program under award numbers 80NSSC23M0191 and 80NSSC23M0192. 

\section*{Author contributions statement}

Y.Y. and Y.C. conceived the study. Y.Y. conducted the data preparation, model development, and experiments. Y.Y. and Y.C. analyzed the results and wrote the manuscript. L.Z., K.W., W.M., and T.G. contributed to the scientific interpretation and revision of the manuscript. All authors reviewed the manuscript.

\section*{Data availability}

The code for this study is available at \url{https://github.com/yuyian/SEP-Prediction-V2}. The data, together with the preprocessing scripts, are archived on Zenodo \url{https://doi.org/10.5281/zenodo.21297635} and are documented in a separate data descriptor \cite{Yu2026SEPPRISM}.

\section*{Additional information}

\textbf{Competing interests} The authors declare no competing interests.

\end{document}